\documentclass[aps,prd,twocolumn,groupedaddress,amsmath,amssymb]{revtex4-1}
\usepackage{graphicx}  
\usepackage{dcolumn}   
\usepackage{bm}        
\usepackage{verbatim}   
\usepackage{hyperref}                   
\def\beq{\begin{equation}}
\def\eeq{\end{equation}}
\def\bea{\begin{eqnarray}}
\def\eea{\end{eqnarray}}

\begin{document}

\title{Thermodynamic Geometry of Normal (Exotic) BTZ Black Hole Regarding to the Fluctuation of Cosmological Constant}
\author {Hosein Mohammadzadeh}
\email{mohammadzadeh@uma.ac.ir}
\affiliation{Department of Physics, University of Mohaghegh Ardabili, P.O. Box 179, Ardabil, Iran}
\author {Maryam Rastkat}
\affiliation{Department of Physics, University of Mohaghegh Ardabili, P.O. Box 179, Ardabil, Iran}
\author {Morteza Nattagh Najafi}
\affiliation{Department of Physics, University of Mohaghegh Ardabili, P.O. Box 179, Ardabil, Iran}


\begin{abstract}
We construct the thermodynamic geometry of $2+1$ dimensional normal (exotic) BTZ black hole regarding to the fluctuation of cosmological constant. We argue that while the thermodynamic geometry of  black hole without fluctuation of cosmological constant is a two dimensional flat spaces, the three dimensional thermodynamics parameters space including the cosmological constant as a fluctuating parameter will be curved. Some consequence of the fluctuation of cosmological constant will be investigated. We show that such fluctuation leads to a thermodynamics curvature which is correctly singular at critical surface of the system. Also, we consider the validity of first thermodynamics law regarding to the fluctuation of cosmological constant.
\end{abstract}
\maketitle

\section{Introduction}\label{1}
Black holes are thermodynamic systems with an entropy related to the area of event horizon by Bekenstein formula, characteristic Hawking radiation related to the surface gravity on the event horizon and internal energy related to the mass \cite{bekenstein1973black,bardeen1973four,hawking1974black}. The entropy of black holes is a function of mass ($M$), angular momentum ($J$) for rotating black holes and charge ($Q$ ) for charged one. Thermodynamics of a large number of black holes such as Schwarzschild , Kerr, kerr-Newman and Reissner-Nordstromm have been studied and interesting results obtained about the stability and probable phase transitions of the system \cite{davies1978thermodynamics,hawking1976black,caldarelli2000thermodynamics,wald2001thermodynamics,chamblin1999holography,cai2004note,peca1999thermodynamics}. One of the most popular approaches for investigation of the thermodynamics of the systems in last decades is the thermodynamic geometry.

One can construct a thermodynamics parameters geometry for any thermodynamics system and define metric elements based on the second derivatives of the entropy with respect to the related extensive thermodynamics fluctuating parameters, which is called the Ruppeiner geometry. Also, the second derivatives of internal energy with respect to the related fluctuating extensive parameters introduce the Weinhold metrics elements \cite{ruppeiner1995riemannian, ruppeiner1979thermodynamics,weinhold1975metric,weinhold1978geometrical}. It has been shown that the curvature of the thermodynamics parameters space, called thermodynamic curvature ($R$) has beneficial information about the systems. The sign of thermodynamic curvature may be used to divide the systems to the intrinsic attractive or repulsive statistical interactions \cite{janyszek1990riemannian}. Also, for an ideal non interacting classical gas the thermodynamics parameters space is flat \cite{nulton1985geometry}. A more special result about the thermodynamics curvature is related to its singular behavior at phase transition points of the system \cite{ruppeiner1995riemannian,janke2004information}.

The thermodynamic curvature of many well-known thermodynamics systems has been considered extensively \cite{ruppeiner1995riemannian}. Recently, the thermodynamics geometry of some generalized statistics have been investigated and their thermodynamical and statistical properties considered \cite{mirza2010thermodynamic,mirza2011thermodynamic,mohammadzadeh2016perturbative,mohammadzadeh2017thermodynamic,talaei2017condensation}. Moreover, thermodynamic geometry of different black holes as a thermodynamics system have been studied by numerous authors \cite{aaman2003geometry,shen2007thermodynamic,ruppeiner2007stability,aaman2008thermodynamic,ruppeiner2008thermodynamic,sarkar2006thermodynamic,sahay2010thermodynamic1,sahay2010thermodynamic2,sahay2010phase,banerjee2011new,banerjee2011second,niu2012critical,wei2012thermodynamic,mansoori2014correspondence,mirza2007ruppeiner,poshteh2013phase,mansoori2016extrinsic}. Aman {\it et al.} investigated the geometry of several black holes thermodynamics in 2003 \cite{aaman2003geometry}. They showed that the geometry of thermodynamics parameters space is flat for the BTZ and Reissner-Nordstrom black holes, while curvature is singular for the Reissner-Nordstrom anti-de Sitter and Kerr black holes. Due to the interacting nature of  constituents of the black holes, the zero curvature of thermodynamic geometry seems unwarranted. Also, a well-known thermodynamics system, the ideal classical gas, with flat geometry has not any phase transition.

B. Mirza and M. Zamaninasab used an extended set of thermodynamics parameters and deduced that the thermodynamic geometry of Reissner-Nordstrom could be non flat \cite{mirza2007ruppeiner}. In fact, they considered the thermodynamic geometry of a usual charged rotating black hole in anti de sitter space, namely Kerr-Newman-AdS black hole.
They evaluated the curvature and then obtained the limit of zero angular momentum and zero cosmological constant, which corresponds to the Reissner-Nordstrom black holes and argued that the curvature is not zero.

There is a $2+1$ dimensional solution for Einstein equation in anti de sitter space. When the cosmological constant is zero, a vacuum solution of $2+1$ dimensional gravity is necessarily flat and it can be shown that no black hole solution with event horizon exist. Negative cosmological constant provides a solution, thereinafter a $2+1$ dimensional Black holes which introduced by M. $\mathrm{Ba\tilde{n}ados}$, C. Teitelboim, and J. Zanelli \cite{banados1992black}. We will explore the thermodynamic geometry of BTZ black hole. We show that the curvature is zero if the cosmological constant be a necessarily  fixed constant. We will select an appropriate set of extended thermodynamics parameters that encompasses the fluctuation of cosmological constant and consider the consequence of such fluctuation. It is well known that in low dimensions the fluctuations can play more important roles. Therefore, fluctuation of cosmological constant in $2+1$ dimension might be more relevant with respect to higher dimensions. We will argue that the extended thermodynamic geometry presents consistence conceptual and physical results. Specially, we can specify the correct phase transition points which correspond to the singular points of curvature of extended thermodynamics parameters space.

Recently, the thermodynamics of black holes has been developed in extended phase space, where the cosmological constant is considered as thermodynamic pressure and treated as thermodynamic fluctuating parameter in its own right. In this approach, the mass of black holes does not mean the internal energy, rather it is modified to chemical enthalpy. َAlso, there exist some good reasons why the fluctuation of cosmological constant should be included in thermodynamic consideration which we will review briefly in the following \cite{kubizvnak2017black,frassino2015lower,kubizvnak2012p}.

The current paper has been organized as follows. In section II we introduce the normal BTZ black holes and work out the thermodynamics curvature. Also, we obtain heat capacities of the system and show that the cosmological constant has to be a fluctuating thermodynamics parameter. Section III will be assigned to the thermodynamic geometry and heat opacities of the exotic BTZ black hole. We consider the validity of first thermodynamics law for normal and exotic black holes regarding the fluctuation of cosmological constant in section IV. Finally, we conclude the paper in section V.
\section{Fluctuation of cosmological constant}
The idea of including the cosmological constant to the dynamical variables or a fluctuating thermodynamics parameters was proposed by Teitelboim and
Brown \cite{brown1988neutralization} and the corresponding thermodynamic term was formally incorporated into
the first law of thermodynamics somewhat later \cite{caldarelli2000thermodynamics}, though no interpretation of the conjugate variable was considered. By permitting the cosmological constant to be a variable quantity or fluctuating thermodynamics parameter we recover the familiar pressure-volume
term from chemical thermodynamics \cite{kubizvnak2017black}. 

Recently, investigation of the first law of black hole thermodynamics by taking into account the fluctuation of cosmological constant, $\Lambda$, has attracted keen interest from researchers \cite{caldarelli2000thermodynamics,kubizvnak2017black,frassino2015lower,kubizvnak2012p,cvetivc2011black}.
Fluctuation of $\Lambda$ in the first law of thermodynamics of a black hole corresponds to considering black hole ensembles with different asymptotics. Three good arguments can be found in the literature to including the variation of cosmological constant in thermodynamic considerations. First, the physical constants such as Yukawa coupling, gauge coupling constants, Newton's constant and cosmological constant can be variable in more fundamental theories and arise as vacuum expectation values. Therefore, the variation of such constants should be included to the first law \cite{gibbons1996moduli,creighton1995quasilocal}. Second, The first law of black hole thermodynamics is inconsistent with the Smarr-Bekenstein relation in the presence of a cosmological constant and we will illustrate it in the following \cite{kastor2009enthalpy}. It has been shown that for Born-Infeld black holes, the maximal field strength $b$ should be included to the fluctuating thermodynamics parameters to obtain a consistent corresponding Smarr relation \cite{breton2005smarr,yi2010energy}. In the same manner one can consider the cosmological constant as a fluctuating thermodynamics parameter. Third, including the cosmological constant to fluctuating thermodynamics parameter in the first law, changes the internal energy interpretation of black hole mass, $M$, to the enthalpy. Since the fluctuating cosmological constant corresponds to pressure, its thermodynamical conjugate has dimension of volume which called thermodynamics volume \cite{kastor2009enthalpy}.

The black hole chemistry has been introduced in extended thermodynamics parameters space regarding to the cosmological constant of $\mathrm{AdS}$ spacetime as a fluctuating parameter, similar to pressure in the first law of thermodynamics \cite{frassino2015lower,creighton1995quasilocal,kastor2009enthalpy,dolan2011compressibility,kubizvnak2012p}. From the perspective of cosmology, a negative cosmological constant induces a vacuum pressure. As we mentioned above, the black hole mass plays the role of gravitational version of chemical enthalpy which is the total energy of black hole, including both the internal energy and the energy required to displace its environment. In fact for a spacetime with fluctuating negative cosmological constant, a new term is added to the first law, which can be regarded as a displacement of vacuum energy \cite{frassino2015lower}.

$\mathrm{AdS/CFT}$ correspondence, which relates a (quantum) gravitational theory in $d$-dimensional $\mathrm{AdS}$ space to a conformal field theory on its $(d-1)$-dimensional boundary motivates one to study the $\mathrm{AdS}$ black holes. According to this correspondence, the cosmological constant is fixed. Therefore, the thermodynamic pressure related to the fluctuating $\Lambda$ and the thermodynamic volume of a black hole have different interpretation on the boundary $\mathrm{CFT}$. Thermodynamic pressure related to the fluctuation of $\Lambda$ is defined as follows \cite{kubizvnak2017black}
\bea
P=-\frac{\Lambda}{8\pi G_d}=\frac{(d-1)(d-2)}{16\pi l^2 G_d},
\eea
where, $G_d$ is the $d$-dimensional gravitational constant and $AdS$ curvature radius $l$ is a measure the number of degrees of freedom, $N$, of the boundary filed theory. The relation between $l$ and $N$ is specified by the family of $\mathrm{CFT}$s being considered. It has been suggested that the fluctuation of pressure or equivalently $\Lambda$ is equivalent to fluctuation of the number of degrees of freedom, $N$, in the boundary $\mathrm{CFT}$. The thermodynamic conjugate of pressure , thermodynamic volume, should be interpreted as an associated chemical potential in the boundary field theory \cite{johnson2014holographic,dolan2014bose,kastor2014chemical}. Another interpretation suggests that $N$ should be fixed, so that we are always referring to the same field theory. Therefore, fluctuation of $\Lambda$ has the more natural consequence of fluctuation the volume of the field theory. However, to stay at fixed $N$ as the volume is varied, we have to compensate by variation of $G_d$ \cite{karch2015holographic}.

Reconsideration of black hole regarding to the fluctuation of cosmological constant has led to the realization that thermodynamics of black hole is much richer topic than previously thought. Moreover the introduction of pressure, and with it a concept of  thermodynamics volume for a black hole,
new phase behaviour, similar to that seen in gels and polymers, was found to be present. Also, triple points were discovered for black holes, analogous to those in water. More recently, black holes have been investigated as heat engines \cite{kubizvnak2014black,mann2016chemistry,johnson2014holographic,johnson2016gauss,johnson2016born,setare2015polytropic}.
 
\section{Negative cosmological constant and BTZ solution}
Albert Einstein introduced an additional term to his theory of general relativity to hold back gravity and achieve a static universe \cite{einstein1923kosmologische}. This term takes into consideration the value of energy density of the vacuum of space. In this way, the cosmological constant; $\Lambda$, appears in well known Einstein field equation as follows
\bea
R_{\mu\nu}-\frac{1}{2}R g_{\mu\nu}+\Lambda g_{\mu\nu}=\frac{8\pi G}{c^4}T_{\mu\nu},
\eea
where, all terms comply with the known standard definition. In 1992, a black hole solution (BTZ) for $2+1$-dimensional topological gravity with a negative cosmological constant was discovered. On the other hand, one can find the entropy of black holes using the Bekenstein-Hawking relation and construct the thermodynamics of any black holes. What is the role of cosmological constant in thermodynamics of the black holes? Is it possible that the cosmological constant behaves such as a fluctuating thermodynamics parameter? It seems that we can consider the thermodynamics of black holes regarding to the fluctuation of cosmological constant according to the arguments and motivations of previous section.
\subsection{2+1 Dimensional BTZ black hole}
The metric of BTZ black hole in the absence of charge in $2+1$ dimension is given by \cite{banados1992black}
\bea
{ ds^2 = -{\Delta\ dt^2}+{dr^2\over\Delta}+{r^2}{(d\phi}-{J \over 2r^2}\ dt)^2 },
\eea
 where,
 \bea
  {\Delta=-{M} + \frac{r^2}{l^2} + \frac{J^2}{4r^2}},
 \eea
 where, $M$ and $J$  are mass and angular momentum, respectively and $l$ is related to cosmological constant $\Lambda$ by
 \bea
 {\Lambda=-{1\over l^2}}.
 \eea
 One can obtain the internal and external horizon radius using equation
 \bea
  {\Delta={0}}
 \eea
  and due to Bekenstein-Hawking formula, black hole entropy is attained. Lastly, we can calculate black hole's mass as a function of entropy $S$,  angular momentum $J$ and $l$ as follows
 \bea
  {M={S^2\over 16\, l^2}+{4\,J^2\over S^2}}\label{mass}.
 \eea
 \subsection{Thermodynamic geometry}
 A geometrical approach has been introduced by Ruppeiner and Weinhold for considering the different thermodynamical systems \cite{ruppeiner1979thermodynamics,weinhold1978geometrical}. In fact, one can construct a thermodynamics parameters space and define an appropriate metric for this space. The curvature of thermodynamics parameters space can be evaluated and is called thermodynamic curvature. Thermodynamic curvature is a useful quantity for investigation of thermodynamical properties of the systems.

 The thermodynamic curvature of an ideal classical gas is zero. It means that the thermodynamics parameters space is flat. The curvature of thermodynamics parameters is related to the intrinsic curvature of constituent particles of the system. There is no interaction between particles in an ideal gas and therefore, the curvature is zero. For an ideal boson (fermion) gas the intrinsic statistical interaction is attractive (repulsive) and the curvature is negative (positive). Also, another important application of the thermodynamic curvature is related to the phase transition point. It has been shown that the thermodynamic curvature is singular in phase transition points. Therefore, one can explore the transition points of a thermodynamical system by considering the singular points of the curvature.

 There are different definitions for the metric elements of the thermodynamic parameters space. Ruppeiner metrics is defined by the second derivatives of the entropy of the system with respect to the appropriate extensive fluctuating parameters. Also, Weinhold metrics is constructed by the second derivatives of the internal energy of the system with respect to the other appropriate extensive fluctuating parameters. It is straightforward to show that these metrics are the conformal transformation of each other. There are also other types of the metrics in the literature \cite{crooks2007measuring}. we focus on the Ruppeiner definition.

 Now, we construct the geometry of thermodynamic parameters of BTZ black hole. Eq. (\ref{mass}) indicates that the mass of black hole which is equivalent to the internal energy of the system is a function of two fluctuating parameters $S$ and $J$ and a constant parameter $l$. Therefore, the thermodynamics parameters space is two dimensional and we could obtain Ruppeiner metric elements according to the definition of the metric in the following equation
\bea
g_{ij}=\frac{1}{T}\frac{\partial^2 M}{\partial X^i X^j},~~~~~~~~~~~~~{X^{i}=(S,J)},
\eea
where,
\bea
 {T={\partial M \over \partial S}}.
 \eea
 Therefore, all metric elements are obtained as follows
 \bea
 g_{SS}&=&-{{192\,J^2 l^2 +S^4}\over {S(64\, J^2l^2-S^4)}},\nonumber\\
 g_{SJ}&=&g_{JS}={{128\, Jl^2}\over {64 \,J^2l^2-S^4}},\\
 g_{JJ}&=&-{{64\, Sl^2}\over {64\, J^2l^2-S^4}}.\label{elements2}\nonumber
 \eea
 Evaluation of thermodynamic curvature is now straightforward. We obtain a simple but unjustifiable result,
 \bea
 R=0.
 \eea
This is a contradicting result. The zero curvature is a symptom of non interacting particles of the system. For a black hole an attractive gravitational interaction is expected. The nature of attractive gravitational interaction is inconsistent with zero thermodynamic curvature of the system.
\subsection{Heat Capacity}
\begin{figure}
\centerline{\includegraphics[scale=0.4]{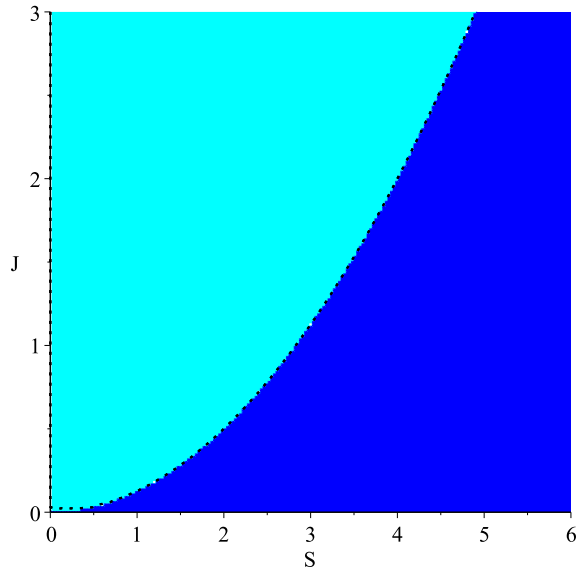}}
\caption{Region positive heat capacity (dark blue) and negative heat capacity (light blue). The dotted line ($C_{Jl}=0$) separates stable and non-stable phase. For simplicity we set $l=1$.}
\label{Fig1}
\end{figure}

Important aspects of the thermodynamics systems such as black holes need to know the heat capacities $C_{x}$, where ${x}$ denotes the set of thermodynamics parameters held constant. The most important information are obtained from response functions, such as heat capacities which is given by
\bea
C_{x}=({\partial M \over \partial T})_{x}=T ({\partial S \over \partial T})_{x}
\eea
Using Eq. (\ref{mass}), we can work out the heat capacity at fixed angular momentum while the parameter $l$ is supposed to be a constant.
\bea
C_{Jl}=({\partial M \over \partial T})_{Jl}=\frac{({\partial M \over \partial S})_{Jl}}{({\partial T \over \partial S})_{Jl}}
={\frac { \left( {S}^{4}-64\,{J}^{2}{l}^{2} \right) S}{192\,{J}^{2}{l}
^{2}+{S}^{4}}}
\eea
The thermodynamic conjugate of angular momentum, namely angular velocity is defined by
\bea
{\Omega={\left(\frac{\partial M}{\partial J}\right)}_{Sl}},
\eea
and the other response function, heat capacity at fixed angular velocity is obtained as follows
\bea
C_{\Omega l}=({\partial M \over \partial T})_{\Omega l}=\frac{({\partial M \over \partial S})_{\Omega l}}{({\partial T \over \partial S})_{\Omega l}}
={\frac { \left( 64\,{J}^{2}{l}^{2}+{S}^{4} \right) S}{{S}^{4}-64\,{J}^{2}{l}^
{2}}},
\eea
where, Nambo brackets definition was applied for the last step. \cite{mansoori2015hessian}. We have specified two regions of stable (positive capacities) and non stable (negative capacities) in Fig. (\ref{Fig1}). The curve between two regions, where $C_{jl}=0$ and $C_{\Omega l}$ diverges is the critical curve and specifies a phase transition from stable to non stable black holes. However, the evaluated zero thermodynamics curvature does not contain any information about the phase transition. In fact, there is a contradiction between zero thermodynamic curvature and behavior of heat capacities of the systems. What is the origin of this contradiction?
\subsection{Fluctuating cosmological constant}
\begin{figure}
\centerline{\includegraphics[scale=0.4]{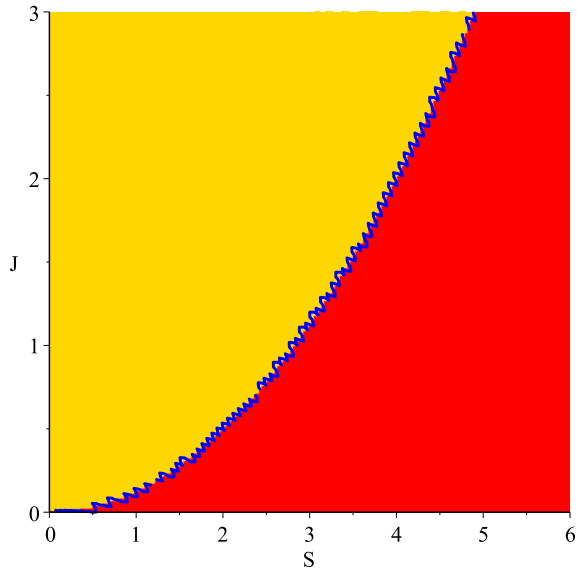}}
\caption{positive thermodynamics curvature (dark/red) region and negative thermodynamics curvature (light/golden) region. The wavy line separates stable ($R>0$) and non-stable ($R<0$) phase and the curvature is singular along this critical line. For simplicity we set $l=1$.}
\label{Fig2}
\end{figure}

Exploring the origin of above contradictions leads us to reconsider the constructed geometry of thermodynamics parameters. We supposed that the thermodynamics parameters is a two dimensional space which is constructed by the fluctuating parameters $S$ and $J$, while $l$ is a constant. Now, we consider $l$ as a fluctuating thermodynamics parameter. Therefore, the thermodynamics parameters space will be a three dimensional space. The elements of Ruppeiner metrics can be evaluated by
\bea
 g_{ij}=\frac{1}{T}\frac{\partial^2 M}{\partial X^i X^j},~~~~~~~~~~~~~{X^{i}=(S,J,l)}.
\eea
Four elements of metrics defined in Eqs. (\ref{elements2}). The other metric elements are given by
\bea
&&g_{Sl}=g_{lS}={{2 \,S^4}\over {l(64\, J^2l^2-S^4)}},\nonumber\\
&&g_{Jl}=g_{lJ}=0,\\
&&g_{ll}=-{{3\, S^5}\over {l^2(64\, J^2l^2-S^4)}}.\nonumber
\eea
It is straightforward to work out the thermodynamics curvature using the evaluated metrics elements  similar to the previous section.
\bea
{R=-{1\over 2}\,{{86016\,J^4l^4 + 3456 \,J^2l^2S^4 + 5 \,S^8}\over {S(192 \,J^2 l^2 +S^4)(64\, J^2l^2-S^4)}}}
\eea
Fig. (\ref{Fig2}) shows that there are two distinct regions similar to the two distinct regions in Fig. (\ref{Fig1}). Positive (negative) curvature region is coinciding with stable (non stable) phase where the heat capacities are positive (negative). In fact, the thermodynamic curvature is positive (negative) where the heat capacities are positive (negative). The system is interactive and the thermodynamics parameters space in not flat. Also, the heat capacities predict phase transition along a critical curve where the thermodynamic curvature is singular.
\section{Exotic BTZ black holes}
\begin{figure}
\centerline{\includegraphics[scale=0.4]{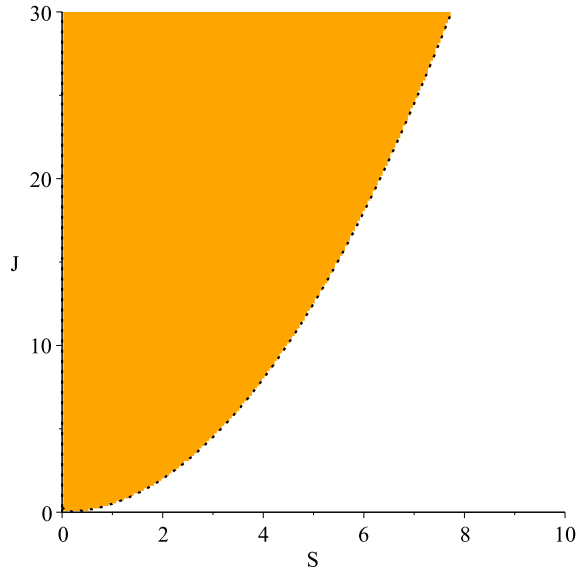}}
\caption{Real valued region for mass. The outer region is not physical. For simplicity we stetted $l=1$.}
\label{Fig3}
\end{figure}

In the exotic Einstein gravity one can find a BTZ metric solution \cite{horne1989conformal,afshar2012conformal}. The metric is the same as that of three dimensional Einstein gravity but reversed role for angular momentum and mass \cite{park2007thermodynamics,townsend2013thermodynamics}. The other difference is related to the entropy. The entropy is given by the length of the inner horizon instead of the event horizon for normal BTZ solution. Such black holes have been called exotic. The mass ($M_E$) of exotic BTZ black holes has been derived as a function of entropy ($S$), angular momentum ($J$) and the cosmological constant ($l$) as follows \cite{zhang2016mass,liang2017smarr}
\bea
M_{E}=\sqrt{\frac{2 {S}^{2}J}{l^{3}}-\frac{{S}^{4}}{l^{4}}}.\label{massE}
\eea
It has been shown that the temperature has negative value, due to an upper bound of mass as in rotating sysyem, and the angular velocity has a lower bound. It is obvious from Eq. (\ref{massE}) for fixed value of $l$, there is a lower bound on the value of angular momentum for a specified value of entropy. Fig. (\ref{Fig3}) shows the region where the mass ($M_{E}$) has real physical value.
We can evaluate the temperature as follows
\bea
T=\frac{\partial M_{E}}{\partial S}={\frac {2(Jl-{S}^{2})}{{l}^{2}\sqrt {2\,Jl-{S}^{2}}}}.
\eea
\begin{figure}
\centerline{\includegraphics[scale=0.4]{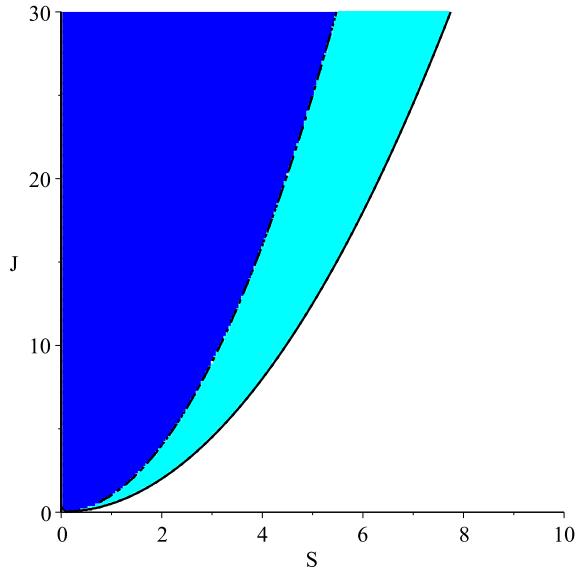}}
\caption{Region with positive temperature (dark blue) and negative temperature (light blue) in physical range. For simplicity we stetted $l=1$.}
\label{Fig4}
\end{figure}
Fig. (\ref{Fig4}) shows that in physical range, the temperature could be negative and positive. We consider in the following the thermodynamic geometry of exotic BTZ black holes in physical range for two cases of negative and positive temperature.
\subsection{Thermodynamic geometry}
We introduced the Ruppeiner metric in previous section. First, we suppos that the cosmological constant and therefore, $l$ is a non fluctuating fixed parameter. Therefore, the thermodynamics parameters construct a two dimensional space and the metric elements are given by
\bea\label{emetricE}
&&g_{SS}=-{\frac {S \left( 3\,Jl-{S}^{2} \right) }{ \left( 2\,Jl-{S}^{2} \right)  \left( Jl-{S}^{2} \right) }},\nonumber\\
&&g_{SJ}=g_{JS}={\frac {{l}^{2}J}{ \left( 2\,Jl-{S}^{2} \right)  \left( Jl-{S}^{2} \right) }},\\
&&g_{JJ}=-{\frac {S{l}^{2}}{2 \left( 2\,Jl-{S}^{2} \right)  \left( Jl-{S}^{2} \right) }}.\nonumber
\eea
It is straightforward to evaluate the thermodynamic curvature using Eqs. (\ref{emetricE}) and find that the two dimensional thermodynamics parameters space is flat for exotic BTZ black holes similar to normal BTZ black hole. Again, we deal with a conceptual problem such as normal BTZ black hole. Black holes are thermodynamic systems with interacting gravitating constituents while the zero thermodynamic curvature is related to a non interacting system such as classical ideal gas.
\subsection{Heat Capacity}
Similar to the previous section, we work out the heat capacity and investigate the probable phase transition for exotic BTZ black hole. Similar to previous section we extract two heat capacity as follows
\bea
&&C_{Jl}={\frac { \left( {S}^{2}-2\,Jl \right)  \left( {S}^{2}-Jl \right) }{S
 \left( {S}^{2}-3\,Jl \right) }},\label{capacity1}\\
&& C_{\omega l}={\frac { \left({S}^{2}-2\,Jl \right) S}{{S}^{2}-Jl}}.\label{capacity2}
\eea
\begin{figure}
\centerline{\includegraphics[scale=0.4]{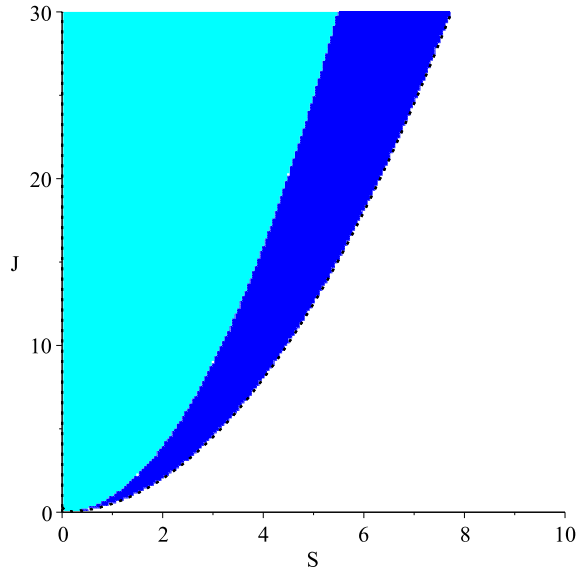}}
\caption{Region with positive heat capasities $C_{Jl}$ and $C_{\omega l}$ (light blue) and negative Heat capacities (light blue) in physical range. For simplicity we stetted $l=1$.}
\label{Fig5}
\end{figure}
We have identified two regions in Fig. (\ref{Fig5}). A comparsion of Fig. (\ref{Fig5}) and Fig. Fig. (\ref{Fig4}) shows that the heat capacities are positive in negative temperature region and conversely the heat capacities are negative in positive temperature region. The intersection of these two regions could be interpreted as the phase transition line. Acrosa this line $C_{Jl}=0$ and the system stability  is changed. Also this line coincides with the singularity of $C_{\omega l}$.

We notice that according to Eq. (\ref{capacity1}), the sign of $C_{Jl}$ changes across two surfaces in parameter space; $ \left( {S}^{2}-2\,Jl \right)=0$ and $ \left( {S}^{2}-Jl \right)=0$; and $C_{Jl}$ diverges across the surface $\left( {S}^{2}-3\,Jl \right)=0$. However, the only phase transition surface is $ \left( {S}^{2}-Jl \right)=0$, located on the physical range and two other surfaces are out of the physical range of the parameters. A similar argument is valid for the $C_{\omega l}$.

We argue that on the one hand the interactional nature of black holes and on the other hand, the existence of phase transition in the system is inconsistent with the zero curvature of thermodynamic parameters space. As a remedy, we consider the cosmological constant as a fluctuating thermodynamic parameter.
\subsection{Again, fluctuating cosmological constant}
By including the $l$ in the set of fluctuating parameters, we deal with a three dimensional thermodynamic parameters space. The thermodynamic metrics of the system has nine elements and four elements has been defined in Eqs. (\ref{emetricE}). The remaining elements are defined as follows
\bea
&&g_{Sl}=g_{lS}=-{\frac {3\,{J}^{2}{l}^{2}-6\,{S}^{2}Jl+2\,{S}^{4}}{l \left( 2\,Jl-{S}
^{2} \right)  \left( Jl-{S}^{2} \right) }}
,\nonumber\\
&&g_{Jl}=g_{lJ}={\frac {S \left( {S}^{2}-3\,Jl \right) }{2 \left( 2\,Jl-{S}^{2}
 \right)  \left( Jl-{S}^{2} \right) }}
,\\
&&g_{ll}={\frac {S \left( 15\,{J}^{2}{l}^{2}-20\,{S}^{2}Jl+6\,{S}^{4}
 \right) }{2{l}^{2} \left( 2\,Jl-{S}^{2} \right)  \left( Jl-{S}^{2}
 \right) }}
.\nonumber
\eea
Using the metric elements, we evaluate the curvature of three dimensional thermodynamics parameters space as follows
\bea
R={\frac {12\,{J}^{3}{l}^{3}+33\,{J}^{2}{S}^{2}{l}^{2}-30\,J{S}^{4}
l+5\,{S}^{6}}{2S \left( 3\,Jl-{S}^{2} \right)  \left( 2\,Jl-{S}^{2}
 \right)  \left( Jl-{S}^{2} \right) }}.
\eea

\begin{figure}
\centerline{\includegraphics[scale=0.4]{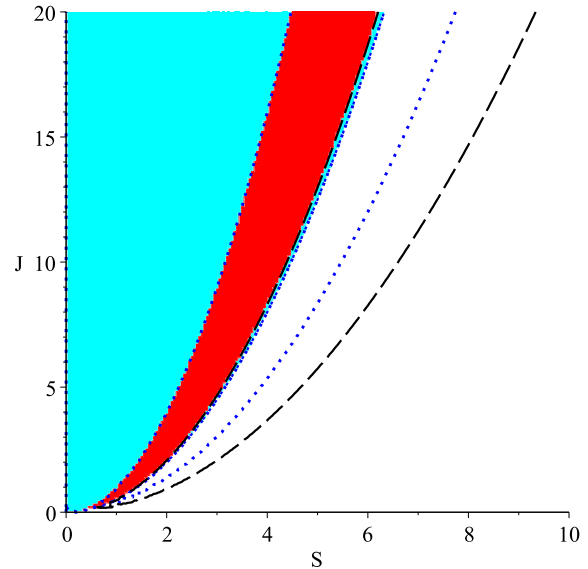}}
\caption{Regions withe positive (light/blue) and negative (dark/red)thermodynamic curvature. The colorless region shows the non physical range. Three doted line are correspond with the singular points of the curvature. It is obvious that only the real phase transition line lies in the physical range.
The dashed lines are corresponds to the zero curvature where the sign of curvature is changed. Only one of these lines is in physical range. For simplicity we stetted $l=1$.}
\label{Fig6}
\end{figure}
Fig.  (\ref{Fig6}) shows that there is a real phase transition surface in the physical range of thermodynamics parameters space; $ \left( {S}^{2}-Jl \right)=0$. This surface coincides with the singular points of the thermodynamics curvature, wherein the temperature of the exotic black holes is zero. In other words, in positive curvature region, the temperature is positive. The zero temperature surface coincide with the singular points of thermodynamic curvature and $C_{\omega l}$ and also, with the $C_{Jl}=0$.
Also, the negative temperature regions are divided to the two subregions with opposite sign of thermodynamic curvature. One can observe that the thermodynamic curvature is negative in negative temperature region except in a narrow region near to the margin of the physical range.

Again, including  $l$ in the fluctuating thermodynamics parameters presents a correct description for the probable phase transitions and the intrinsic statistical interaction of the systems. Also, some significant results about the system can be explored by fluctuation of the cosmological constant.
\section{First Law of Thermodynamics}
 All thermodynamics laws for black holes as thermodynamic systems or equivalently four laws of black holes mechanics were previously discovered by B. Carter, A. Hawking and J. Bardeen \cite{bardeen1973four}. It has been shown that the differential form of the first law of thermodynamics for black holes, which is asymptotically flat in each dimension is given by
 \bea
 {dM=TdS+\Omega dJ}
 \eea
 and Bekenstein-Smarr corresponding mass equation is \cite{liang2017smarr,wei2009understanding}
 \bea\label{BS}
 {{D-3\over D-2}\,M=T S+\Omega J},
  \eea
 where, $\Omega$ and $T$ have been defined in pervious section. Obviously in $2+1$ dimension the left side of the equation is equivalent to zero.
\subsection{Normal BTZ Black hole}
 Although the first side of the Eq. (\ref{BS}) is equivalent to zero for $2+1$ dimensional normal BTZ black hole, a direct calculation shows that the second side of the equation is not, so equality is not established. In fact, the second side of the equation for normal BTZ black hole is equivalent to
 \bea\label{TS}
  TS+\Omega J=\frac{1}{8}\frac{S^2}{l^2}.
  \eea
Now, we consider the cosmological constant; more precisely $l$ as a fluctuating thermodynamics parameter. Therefore, the first law and generalized Bekenstein-Smarr equation will be as follows:
 \bea
 {dM=T\,dS+\Omega \, dJ+\Theta \, dl}
 \eea
 \bea\label{GBS}
 {{D-3\over D-2}\,M=T\, S+\Omega\, J+{1\over D-2}\Theta\, l}
 \eea
where,
\bea
{\Theta=({\partial M\over \partial l})_{\tiny{SJ}}}
\eea
is the conjugate variable of fluctuating thermodynamics parameter $l$. A simple calculation shows that the last term in equation (\ref{GBS}) is given by
\bea\label{ll}
 {1\over D-2}\Theta\, l=-\frac{1}{8}\frac{S^2}{l^2},
\eea
According to Eqs. (\ref{TS}) and (\ref{ll}), Eq. (\ref{GBS}) will be satisfied. Therefore, the fluctuation of $l$ as a thermodynamics parameter has an authentic backing and valid reason.
\subsection{Exotic BTZ Black hole}
As in the same manner, we explore the validity of generalized Bekenstein-Smarr equation for the exotic BTZ black hole \cite{zhang2016mass}. We work out that
  \bea\label{TSE}
  TS+\Omega J={\frac {S \left( 3\,Jl-2\,{S}^{2} \right) }{{l}^{2}\sqrt {2\,Jl-{S}^{2}}}}.
  \eea
It seems that the first law does not hold. However, we suppose that $l$ is fluctuating thermodynamic parameter and obtain that
\bea
{\Theta=({\partial M\over \partial l})_{\tiny{SJ}}}={\frac {S \left( 2\,{S}^{2}- 3\,Jl \right) }{{l}^{3}\sqrt {2\,Jl-{S}^{2}}}}.
\eea
Therefore, we evaluate a new term similar to the normal BTZ blac holes as follows
\bea\label{llE}
 {1\over D-2}\Theta\, l=-{\frac {S \left( 3\,Jl-2\,{S}^{2} \right) }{{l}^{2}\sqrt {2\,Jl-{S}^{2}}}}.
\eea
Therefore, the validity of the thermodynamics first law is evident. Thus, we argue that for normal and exotic BTZ black holes, considering $l$ as a fluctuating thermodynamic parameter is necessary.
\section{ Conclusion }

In this paper we promoted the cosmological constant; more precisely $l$; to fluctuating thermodynamics parameters and  investigated its consequences. It seems satisfactory,
and also consistent with interacting nature of a black holes and possible phase transition. We focused on the two $2+1$ dimensional black holes; normal and exotic black holes; and worked out the thermodynamic curvature. We observed that the two dimensional thermodynamics geometry with fluctuating parameters $(S,J)$ is flat in both cases, while the three dimensional thermodynamics geometry with fluctuating parameters $(S,J,l)$ had non-trivial curvature.

Evaluation of heat capacities at fixed angular momentum and fixed angular velocity shows that one can find a sign change and a singularity for $C_{Jl}$ and $C_{\omega l}$, respectively across a critical surface in thermodynamic parameters space of normal (exotic) BTZ black hole. Considering $l$ as a fluctuating thermodynamics parameter shows that the singularity of the thermodynamic curvature coincides with the critical surface.

Finally, we considered the validity of first thermodynamics law for both normal and exotic BTZ black holes. Again, we argued that fluctuation of $l$ as a thermodynamics parameter is necessary to satisfy the first thermodynamics law. Therefore, the consequence of fluctuation of $l$ include the non zero thermodynamics curvature emphasizes on the interacting identity of black holes, singular behavior of thermodynamic curvature on the critical surface and the satisfaction of the first law of thermodynamics.

\acknowledgments
H. M. Would like to thank Professor R. B. Mann and an unknown referee for their useful comments and introduction of some useful papers.

 M. R. would like to express her very great appreciation to Farzad Farrokhi  for his valuable and constructive suggestions about editing this paper. His willingness to give his time so generously has been very much appreciated.

\bibliography{refs}

\end{document}